\documentclass{article}

\usepackage{amssymb}

\usepackage{amsfonts,amsmath,amsthm}
\usepackage{graphicx,epstopdf}
\usepackage{graphicx}
\usepackage{float}

\usepackage{listings}
\usepackage{color}

\usepackage{float}
\usepackage{amsmath} 
\usepackage{calc}
\newlength{\depthofsumsign}
\usepackage{calc}

\begin{document}



\newcommand{\Eqref}[1]{(\ref{#1})}


\begin{center}
	\large{ \textbf{ {\Large An RBF-PSO Based Approach for Modeling Prostate Cancer}}}
\end{center}

\begin{center}
 Emma Perracchione$^*$, Ilaria Stura$^+$
\end{center}

\begin{center}
	$^*$Department of Mathematics "G. Peano", University of Turin - Italy\\
	$^+$Department of Neuroscience,  University of Turin - Italy
\end{center}
\vskip 0.5cm

\textbf{Abstract.} 	
Prostate cancer is one of the most common cancers in men. It is characterized by a slow growth and  it can be diagnosed in an early stage by observing the Prostate Specific Antigen (PSA). However, a relapse after the primary therapy could arise in $25-30\%$ of cases and different growth characteristics of the new tumor are observed. In order to get a better understanding of the phenomenon, a mathematical model involving several parameters is considered. To estimate the values of the parameters identifying the disease risk level a novel approach, based on combining Particle Swarm Optimization (PSO) with a meshfree interpolation method, is proposed.

\section{Introduction}
A universal growth law, the so-called Phenomenological Universalities (PUN) approach, which can be applied also to model the growth of a tumor,
has been developed by \cite{Guiot,Gliozzi2}. In the last mentioned papers, it is shown that the  tumor follows a universal growth law governed  by the  physical meaning of the model parameters. Our aim is to validate such model testing it by challenging real data with popular numerical tools. 

One of the most common tumor in the world is the prostate cancer, which affects mainly men over sixty years old. Luckily, this disease has a very slow growth and a reliable biomarker for the early diagnosis, the Prostate Specific Antigen (PSA). Radical prostatectomy and radio therapy  are the most used primary therapies. After prostatectomy, a relation between PSA and the size of the tumor (metastasis or local relapse) can be reliably observed. 

In this paper, clinical data (Eureka1 data set, \cite{Eureka1}) of prostatectomized patients, collected in Regione Piemonte (Italy) during the last 10 years, are used to investigate the evolution of the relapse of the cancer using PSA series.

The guidelines of the paper are as follows. Section \ref{secMath} is devoted to the presentation of the  mathematical model. 
Then  the \textsc{Matlab}-implemented numerical tools, used to validate the model, are described. Specifically, in Section \ref{secRBF}  the scattered data interpolation problem, solved with the use of  Radial Basis functions (RBFs) is detailed. Section \ref{secPSO} briefly reviews the PSO method. Finally, Section \ref{secResults} and \ref{secConclusions} deal with results and conclusions, respectively.

\section{The mathematical model}\label{secMath}

In  \cite{Guiot} it has been shown how  different existing growth models can be found by truncating at a certain order $n$ the PUN growth law. In particular, when $n=0$ the PUN reduces to the Malthus law, i.e. the tumor grows in the early stage without constraints, while if $n=1$ it becomes the Gompertz law, which successfully models the growth with physical and/or nutrients constraints. Finally, when $n=2$, PUN reduces to the West law. The latter, since it is strictly related to the metabolic rate of the tumor cells, turns out to be a reliable mathematical model for a kind of growth characterized by constraints and metastasis. 

Data of patients who relapsed after prostatectomy are available. To model such data we choose the Gompertzian growth model, characterized by an initial exponential growth, a progressive velocity decrease and finally the achievement of the carrying capacity, due to physical barrier (i.e. organ tissues or membranes) and/or to lack of nutrients. Thus the model reads as follows:
\begin{equation}\nonumber
\frac{dN(t)}{dt}=c_0e^{\beta t}N(t),
\end{equation}
whose solution is:
\begin{equation}\label{eqGompertz}
N(t)=N_0 e^{\frac{c_0}{\beta} (e^{\beta t}-1)},
\end{equation}
where $c_0$ is the growth rate and $\beta$ is somehow inversely proportional to the carrying capacity (see \cite{SturaFixed} for details). 

We expect that, when an early relapse (less than $2$ years after the surgery) or a late relapse (more than $2$ years after surgery) occurs, the value $c_0$ significantly changes.
Similar considerations also hold for the parameter $\beta$ which is related to the aggressiveness of the tumor, i.e. the maximum level which can be reached by the cancer. 
In fact, comparing the patients who have been successfully treated after the relapse and the ones who had other relapses, a different value of $\beta$ is expected. Thus it follows the importance of having an accurate estimation of the parameters, shown in next sections.
%
%
\section{Radial Basis Function}\label{secRBF}

RBF interpolation is a popular numerical tool used to solve the scattered data fitting problem, \cite{Buhmann03,Cavoretto15,Fasshauer}. The method, described in what follows, works in any dimension $d$ but, since in this context it will be  used to reconstruct  a curve, we report here the basic numerical tools concerning scattered data interpolation in the simplest case in which $d=1$. 

Given a set measurements $ {\cal Y}_M= \{ y_i  ,i=1, \ldots, M \}$ and the locations at which these measurements are obtained $ { \cal X}_M= \{ x_i \in \mathbb{R},i=1, \ldots, M \}$, where  $ { \cal X}_M$ is a set of distinct data points or nodes, arbitrarily distributed in a domain $ \Omega \subseteq \mathbb{R}$, the aim is to find a function ${\cal I}: \Omega \longrightarrow \mathbb{R}$ so that ${\cal I}( x_i)=y_i, $ $ i=1, \ldots, M$. Thus the interpolant assumes the form:
\begin{equation}
{\cal I}( x)= \sum_{k=1}^{M} c_k \phi (| x - x_k|), \quad  x \in \Omega,
\label{rad1}
\end{equation}
where $ \phi: [0, \infty) \longrightarrow \mathbb{R}$ is called RBF.

The coefficients $ \{ c_k \}_{k=1}^{M} $ are determined by imposing the interpolation conditions, consequently  the problem leads to the linear system:
\begin{equation}
\Phi c= y ,
\label{sys1}
\end{equation}
where $\Phi \in  \mathbb{R}^{M \times M} $ is:
\begin{equation}
\Phi_{ik}= \phi (|x_i-x_k|), \quad i,k=1, \ldots, M,
\label{A}
\end{equation}
$  c= [c_1, \ldots, c_M]^T$ and 
$  y =[y_1, \ldots , y_M]^T$.

A solution of the problem exists and is unique in the interpolation space if and only if the matrix $\Phi$ is non-singular. 
Therefore, if $ \phi$ is strictly positive definite, the interpolant (\ref{rad1}) is unique, since the corresponding interpolation matrix (\ref{A}) is positive definite and hence non-singular.

Among several examples of strictly (conditionally) positive definite RBFs which can be used to solve the scattered data interpolation problem,  we consider  the 
Mat\'ern $C^2$ function given by:
\begin{equation}
\phi(r)  =  e^{- \varepsilon r} (1+\varepsilon r),
\end{equation}
where $r$ is the Euclidian distance and $\varepsilon$ is the so-called shape parameter determining the smoothness of the function.
%
%
\section{Swarm intelligence} \label{secPSO}

Given a  function $f: \mathbb{R}^d \longrightarrow \mathbb{R}$,  the general statement of the problem consists in finding: 
\begin{equation}
\min_{p} f(p), \quad p \in {\cal D} \subseteq \mathbb{R}^d,
\end{equation}
where ${\cal D}$ is the so-called \emph{feasibility region}, subject to linear or non-linear constraints:
\begin{equation}
g_i(p) \leq 0, \quad  i = 1, \ldots,  m.
\label{constraints}
\end{equation}

To solve such problem we consider common techniques concerning stochastic optimization. The most popular are  evolution strategy (ES) and genetic algorithm (GA), both based on the competition among individuals. On the opposite, we focus on cooperative methods (proposed in the last decades); in particular we choose the particle swarm optimization (PSO), based on the mutual interaction and exchange of information between individuals.

PSO has been first introduced by Kennedy (social psychologist) and Eberhart (electrical engineer), \cite{Kennedy}. Let us consider a group of particles or birds which are represented as points in the space  ${\cal D}$. Once we model the way of fly of the flock, then it results easy to find the minimum of $f$,  taking into account that the target of birds consists in looking for the maximum availability of  food, i.e. the minimum of $f$.

Thus, the main objective is  to simulate trajectories of all single bird by  considering their selfish behavior (which is the ability of a bird of randomly 
fly away from the flock to reach the food) and  the social 	behavior (which is the ability of a bird of stay in the group). With this simple considerations, taking  also into account that particles avoid collisions, it is possible to simulate the way of moving of a group of birds. To explain the idea of how we can find the minimum  of the objective function interpreting the latter as food  which birds look for, let us suppose that a bird discovers some food. Then, if a good trade-off between the two behaviors is reached,  other birds can change their directions towards the same place. Acting in this way, the flock changes gradually its direction until the best place is reached.

In the PSO method, the $n$ particles are randomly initialized in the search-space ${\cal D}$,  with random initial velocities $v^{(1)}_i$.
Then the direction of a single particle and its  velocity gradually change so that it starts to move in the direction of the best previous position of itself or of other birds, searching in a neighborhood a even better position with respect to the fitness measure $f$. In order to define how to update the velocities and the positions of birds, in the loop consisting of $N_{max}$ maximum number of iterations, we remark that the attributes of a single bird, at the $j$-th iteration,  are the current position   $p^{(j)}_i$, the related velocity  $v^{(j)}_i$ and the best position visited by the single bird $l^{(j)}_{i}$, $i=1, \ldots, n$. At the $j$-th iteration the best position among all $l^{(k)}_{i}$, $i=1, \ldots, n$, $k=1, \ldots, j$, is  called the global best position $g^{(j)}$.
Thus it follows that, \cite{Pedersen,Shi}:
\begin{equation}
v^{(j)}_i= \omega^{(j)} v^{(j-1)}_i+ 
\varphi^{(j)}_l (l^{(j-1)}_i-p^{(j-1)}_i)+
\varphi^{(j)}_g (g^{(j-1)}_i-p^{(j-1)}_i), 
\label{up_vel}
\end{equation}
with $i=1, \ldots, n$, and $j=2, \ldots, N_{max}$. Then the position of each bird is computed by adding the velocity to its current position, i.e.:
\begin{equation}
p^{(j)}_i= \omega p^{(j-1)}_i+
v^{(j)}_i. 
\label{up_pos}
\end{equation}
\eqref{up_pos} enforces to the particle to move towards another position, regardless of any improvement to its fitness.

The parameters $\varphi^{(j)}_l$ and $\varphi^{(j)}_g$, involved in the so-called cognitive and social component of \eqref{up_vel}, respectively, are acceleration coefficients and $w^{(j)}$  is the inertia weight.
%
\section{Fitting procedures and results}\label{secResults}
We recall that our aim consists in finding the best values of $c_0$ and $\beta$ for which Equation \eqref{eqGompertz} fits accurately the PSA data set. Therefore, the objective function is:
\begin{equation}
\sum_{i=1}^{n} \left[ N_i-N_0 e^{\frac{c_0}{\beta} (e^{\beta t_i}-1)} \right] ^2,
\label{scarti}
\end{equation}
where $(t_i,N_i), \ i=1,\dots n$ are the real data. Unfortunately, in our case $4 \leq n \leq 15$ and the median value is $6$, thus in order  to improve the performance of the PSO method,  since the cardinality of samples is small,  we first reconstruct  the curve defined by such few samples via RBF interpolation. Acting in this way, after solving the scattered data interpolation problem, we obtain a larger data set $(t_j,N_j), \ j=1,\dots s$, where $s$ is arbitrary large, as input of PSO. 
\begin{figure*}\label{FigOk}
	\centering
	\includegraphics[scale=0.4]{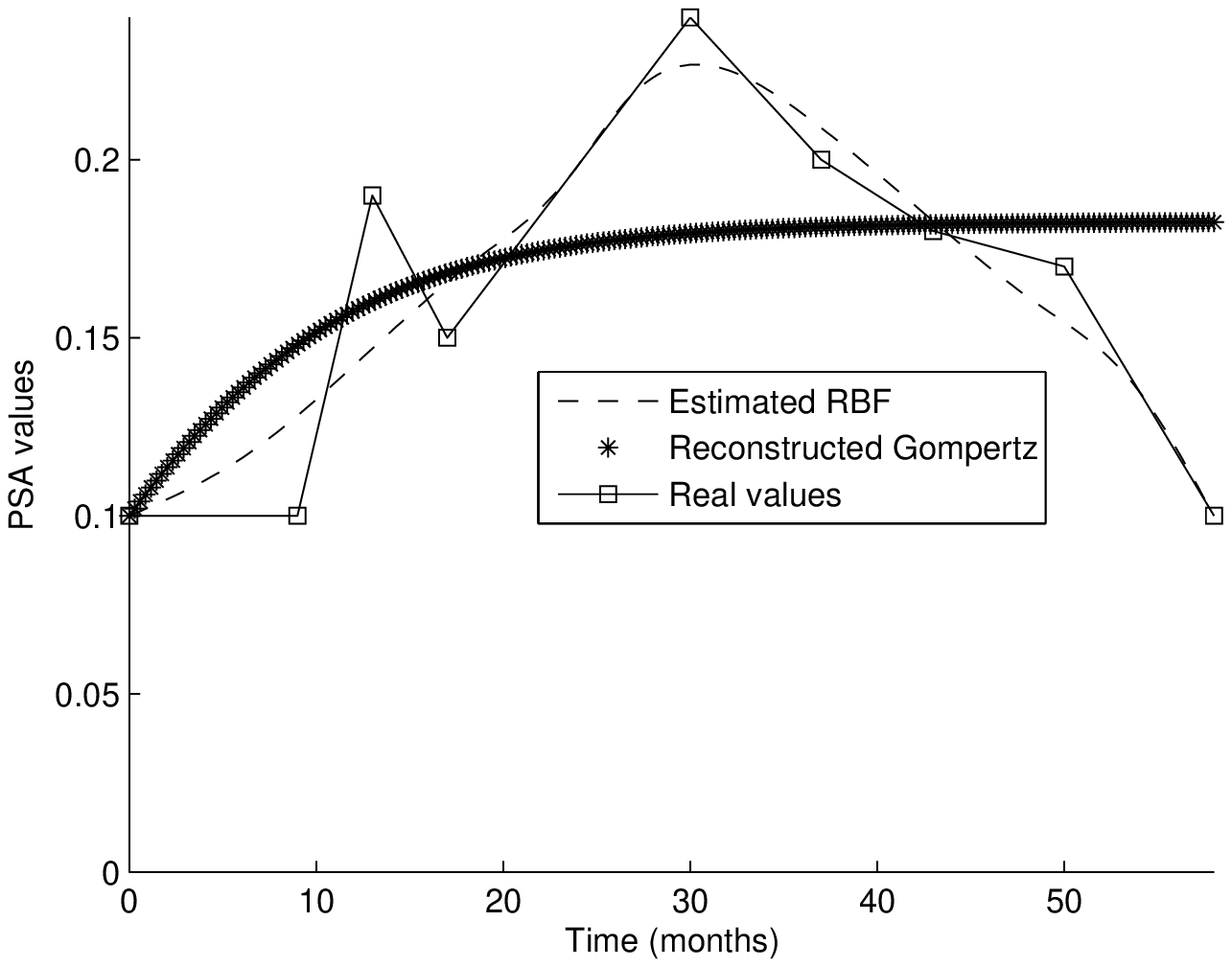}
	\includegraphics[scale=0.4]{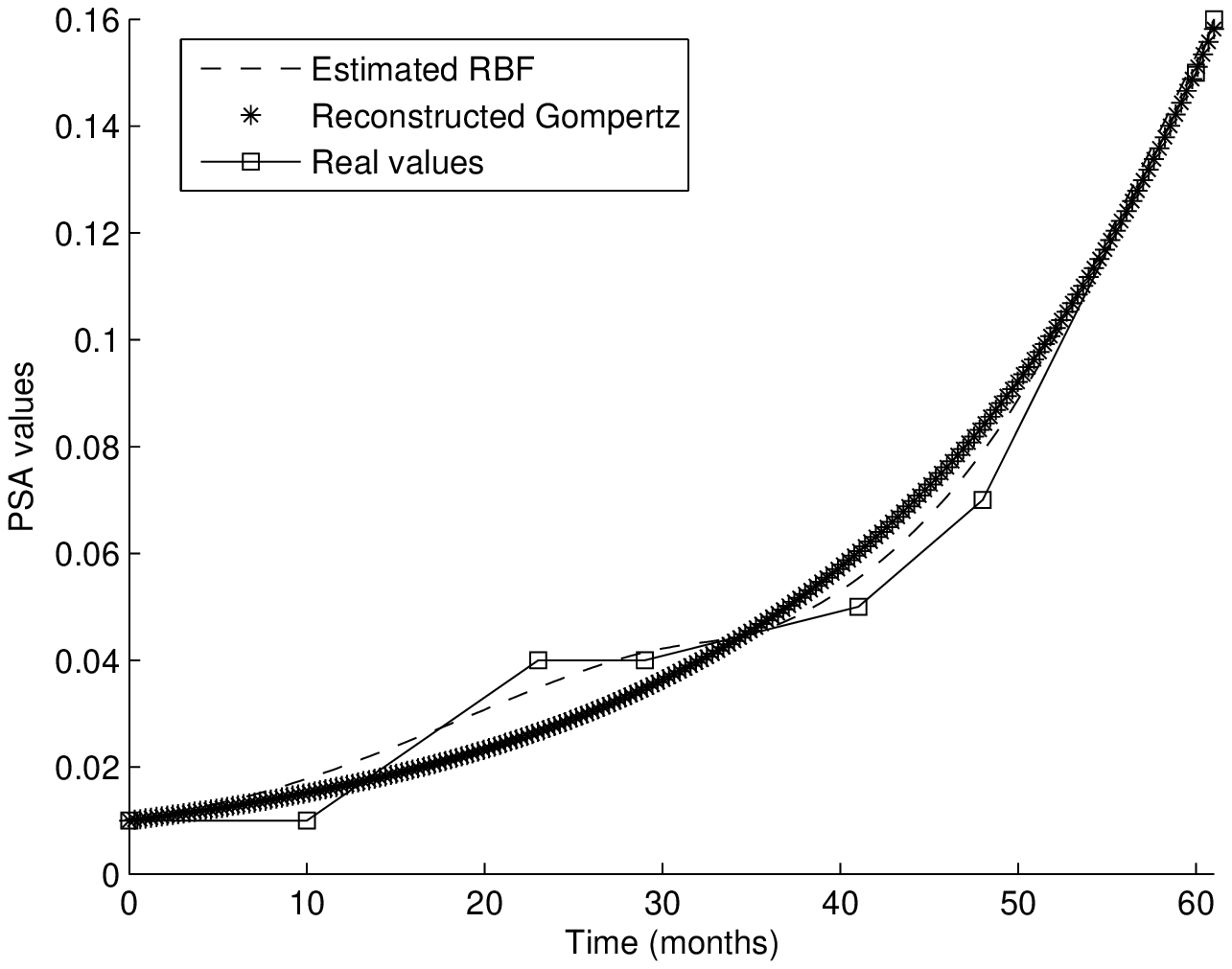}
	\includegraphics[scale=0.4]{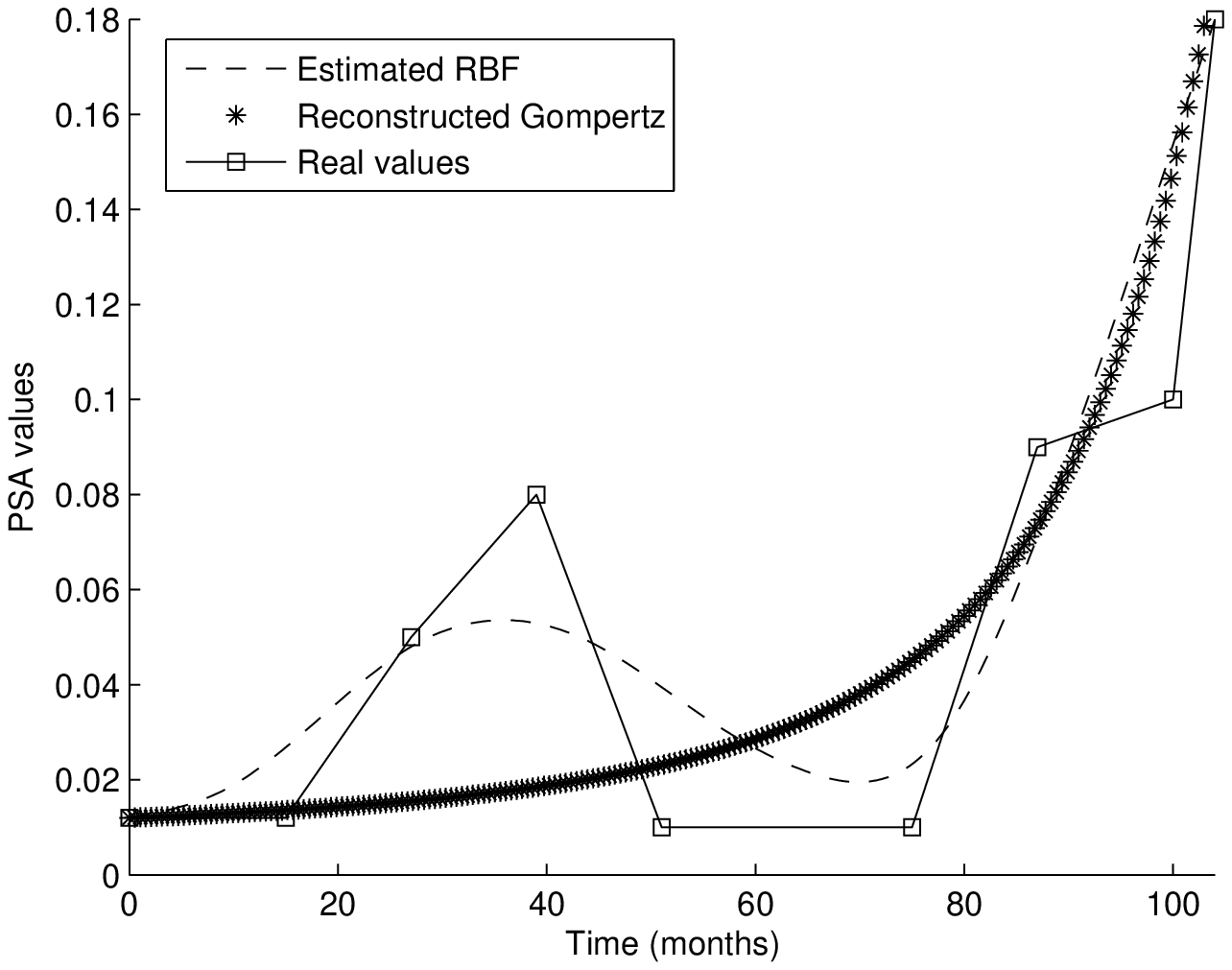}
	\caption{The real data (lines and squares), the RBF interpolant (dotted line) and the gompertzian function (stars) fitted  with parameters estimated by mean of PSO.
		a)  The PSO RBF-based method fails ($c_0=0.007, \beta=-0.117$); a reliable estimation of the parameters is found in b) $c_0=0.04, \beta=0.003$ and c) $c_0=0.007, \beta=0.021$.}
\end{figure*}

We consider $404$ patients of the Eureka1 data set  who relapsed after radical prostatectomy and with at least $4$ PSA available values. 
The $87.4\%$ of PSA data sets, i.e. $353$ patients, can be successfully fitted with the PSO RBF-based method. 
The estimated parameters turn out to be coherent with respect to their physical meaning (see Fig. \label{FigOk}b and \label{FigOk}c).
Moreover, the method we use is reliable and stable since data oscillations do not affect the final result (see Fig. \label{FigOk}c).
As expected, the parameter $c_0$ is larger when the growth is faster. Anyway, regardless of the value assumed by $c_0$,
if $\beta>0.1$ the growth is always very fast and the relapse occurs in the first $24$ months after the surgery ($12$ months if
$\beta>0.2$). Instead, if $\beta$ is very close to $0$, the relapse occurs 
after at least $3$ years.
The parameter $\beta$ does not show obvious correlations with the clinical data (stage of the tumor, presence of lymph nodes,...)
and it does not add useful information in the risk estimation when it is between $0$ and $0.1$. 

The remaining percentage  cannot be successfully fitted with our method. This problem happens when values are decreasing, i.e. when the tumor does not follow the model growth, or if there are three or more identical values in consecutive months,  (see Fig. \label{FigOk}a). Such cases need further investigation.

\section{Conclusions}\label{secConclusions}

In this paper we validate the mathematical model, by considering real data of cancer cell growth.
Specifically, we investigate, by mean of a PSO-RBF-based method, the evolution of the recurrent prostate cancer. We empirically prove that the gompertzian function turns out to be quite an accurate model for the growth of the relapse. Moreover, since we are able to estimate  the growth rate of the tumor given few initial PSA samples, meaningful predictions about the evolution of the cancer can be assessed in the very first period of the relapse. 

In the next future, we want to investigate more deeply the correlation between the parameters values and clinical and genetic data. 
Moreover, a similar investigation with other mathematical models will be helpful to identify the most reliable model for cancer growth. Concerning 
the numerical method, we want to compare the PSO with other optimization algorithms, e.g. the Cuckoo Search and the Bat Algorithm, proposed by \cite{Yang}.

\end{document}